\documentstyle[prd,aps,preprint,tighten,epsfig]{revtex}

\begin{document}
\draft

\title{Small Neutrino Masses From Structural cancellation In Left-Right Symmetric Model}

\author{ {M.J.Luo}\thanks{E-mail: mjluo@mail.ustc.edu.cn}\\
{\sl Department of Modern Physics,\\
University of Science and Technology of China,\\
Hefei, Anhui 230026, China}\\
{Q.Y.Liu}\\
{\sl Department of Modern Physics,\\
University of Science and
Technology of China,\\
Hefei, Anhui 230026, China}}

\maketitle

\begin{abstract}
The Type I, II and hybrid (I+II) seesaw mechanism, which explain why
neutrinos are especially light, are consequences of the left-right
symmetric model (LRSM). They can be classified by the ranges of
parameters of LRSM. We show that a nearly cancellation in
general Type-(I+II) seesaw is more natural than other types of
seesaw in the LRSM if we consider their stability against radiative
correction. In this scenario the small neutrino masses are due to the
structure cancellation, and the masses of the right handed neutrino
can be of order of O(10)TeV. The realistic model for non-zero neutrino masses,
charged lepton masses and lepton tribimaximal mixing can be implemented
by embedding $A_4$ flavor symmetry in the model with perturbations
to the textures.
\end{abstract}


\section{Introduction}
The fact that neutrinos have very small masses has been established by a
number of neutrino oscillation experiments \cite{oscillation} in the
past decade, which is an important evident to go beyond the Standard
Model. In order to generate very tiny neutrino masses, the very
popular explanation is the seesaw mechanism \cite{seesaw}.

In the so-called Type-I seesaw \cite{seesaw}, extra very heavy Majorana right
handed neutrinos (RHN) are introduced. When integrating them out, the
neutrino mass is approximately $m_{\nu} \sim m^2_D/m_R$, so we
assume that the $m_D$ (the neutrino Dirac mass) is of the electroweak
scale, i.e. $m_D \sim O(10^{2}GeV)$, we need the RHN mass $m_R \sim
O(10^{16}GeV)$ which is hopeless to reach to direct test this
mechanism. In the Type-II seesaw (triplet seesaw) \cite{triplet}, a heavy Higgs triplet $\Delta$ is
introduced to play the similar role of heavy right handed neutrino
to suppress the neutrino masses, we have $m_{\nu} \sim
v^2/m_{\Delta}$, where $m_{\Delta} \sim O(10^{16}GeV)$ is the mass of the Higgs triplet. In a
general hybrid Type-(I+II) seesaw model, both terms make contributions to the
neutrino masses. The crucial feature of such mechanisms are
introducing heavy particles to suppress the neutrino masses, but the
smallness of neutrino mass needs them to be too heavy to have any
signals in future colliders.

Possible compromise between the impossible collider signals
of such heavy particles and the smallness of neutrino masses is
discussed in recent literatures in the framework of hybrid Type-(I+II)
seesaw \cite{hybrid}, where the small neutrino masses is from the
structural cancellation, while suppression plays no role. In such
scenario, the introduced heavy particles can be light enough to be
direct produced in future colliders without violating the current
bounds \cite{RHNbound}, so the possibilities have not been ruled out by experimental
limits so far.

These types of seesaw are consequences of the left-right symmetric
model (LRSM) \cite{LRSM}, which is a possible extension of SM. In
the model, unlike the SM that has only $SU(2)$ left handed chiral
matter, the right handed sector under Non-Abelian $SU(2)$
representation are also introduced and correlated to the left handed
sector. The LRSM not only leads to the seesaw mechanism but also
provides explanation of the observed maximal P and C violation at
low energy weak interaction, and is therefore likely in certain
sense to be the final theory. The type of seesaw deduced from LRSM
is determined by the space of parameters of the model. If we
consider the stability of the parameters under the radiative
correction, a model is "natural" if it is stable against the quantum
correction, so fine-tuning for parameters is not needed. Before the
mechanism can be tested directly in experiments, the naturalness is
inevitable an important criteria for our model buildings.

In this paper, we will deduce the three types of seesaw in the LRSM
and classify them by the ranges that the parameters locate. The
1-loop quantum correction of the parameter is evaluated and we find
that the small neutrino mass from nearly cancellation in Type-(I+II)
is more "natural" than other types of seesaw in the limit of small
couplings in Higgs potential. So unlike the literature \cite{hybrid}
where the cancellation relation is imposed by hands, the structure
cancellation in LRSM is a natural result of the model. Therefore, in
this scenario, the RHN can be light and be of order of O(10)TeV.
Finally, non-zero neutrino masses, charged lepton masses and
tribimaximal mixing \cite{TB} are generated by perturbations and
embedding an extra $A_4$ \cite{A4} flavor symmetry into the model.

\section{The Model}
\subsection{The Left-Right Symmetric Model}
The left-right symmetric model is based on the extended gauge group
$G_{LR}=SU(2)_L \otimes SU(2)_R \otimes U(1)_{B-L}$, in which a Higgs
bi-doublet $\Phi$ and left (right) Higgs triplet $\Delta_{L(R)}$ are
introduced and with the representation assignments
\begin{eqnarray}\label{HiggsInG}
\Phi \sim (2,2,0), \, \Delta_L \sim (3,1,2), \, \Delta_R \sim (1,3,2).
\end{eqnarray}
Under a discrete left-right symmetry, $l_L
\leftrightarrow l_R^c$, $\Delta_L \leftrightarrow \Delta_R$ and
$\Phi \leftrightarrow \Phi^T$, the invariant Lagrangian of the Yukawa
interaction term is
\begin{eqnarray}
- {\cal L} & = &  y \overline{l}_{L} \Phi l_R +
\tilde{y}\overline{l}_{L} \tilde{\Phi} l_R +\frac{1}{2}
f[\overline{l}_L i \tau_2 \Delta_L l_L^c +\overline{l^c_R} i
\tau_2 \Delta_R l_R]+ {\rm h.c.},
\end{eqnarray}
where $l_{L(R)}=( \matrix{\nu_{L(R)} & e_{L(R)}})^T$ is the lepton
doublet, $\tilde{\Phi} = \tau_2\Phi^*\tau_2$, $l_{L(R)}^c \equiv C
\overline{l_{L(R)}}^T$ with $C$ being the charge-conjugation matrix.
At first stage, the symmetry spontaneously broken into $SU(2)_L \times
U(1)_Y$ by a non-zero vacuum expectation value (VEV) of $\Delta_R$ , leading to a heavy
Majorana mass for right handed neutrinos. The second stage, the
$\Phi$ develops VEV, breaking the symmetry to relic $U(1)_{em}$.
The developed non-zero VEV consistent with
$U(1)_{em}$ electromagnetic invariance are
\begin{eqnarray}
\langle \Delta_L^{} \rangle = \left( \matrix{0 & 0 \cr v_L & 0 \cr}
\right) \; , \hspace{0.5cm} \langle \Delta_R \rangle = \left(
\matrix{0 & 0 \cr v_R & 0 \cr} \right) \; , \hspace{0.5cm} \langle
\Phi \rangle = \left( \matrix{v & 0 \cr 0 & v' \cr} \right).
\end{eqnarray}
The measurement of the $\rho$ parameter \cite{rho} constrains the
tree-level contribution of the Higgs triplet, $v_L \lesssim 1 GeV$,
which is much smaller than the electroweak scale $v \simeq 174GeV$,
and we will work in the approximation $v' \ll v$. Integrating out
the heavy fields the effective mass of neutrino can be written as
the general Type-(I+II) seesaw formula
\begin{eqnarray}\label{seesaw}
M_{\nu} \simeq M_L -M_D M^{-1}_R M^T_D = v_L f - \frac{v^2}{v_R} y
f^{-1} y^T.
\end{eqnarray}
The dominant contribution from the first or second term determines
the type of seesaw. In the model the charged lepton and Dirac
neutrino mass matrix are simply obtained as $M_e= \tilde{y} v I$ and
$M_D =y v I$ ($I$ is the identity matrix), which we will discuss and
implement by introducing flavor symmetry in section IV.

\subsection{Higgs Potential}
Our aim here is to show the relations between the VEVs of the Higgs
fields in LRSM, for this purpose, let us write the Higgs potential
involving $\Phi$ and $\Delta_{L(R)}$. The most general
renormalizable Higgs fields potential has the quadratic and quartic coupling terms and can not have any trilinear terms. So
consistent with the transformation properties as Eq(\ref{HiggsInG})
and discrete left-right symmetry, the Higgs potential can be written
as \cite{potential}
\begin{eqnarray}\label{v}
V(\Phi,\Delta_L,\Delta_R) &=& -\mu_{ij}^2 tr[\Phi_i^\dagger \Phi_j]
+ \lambda_{ijkl} tr[\Phi_i^\dagger \Phi_j] tr[\Phi_k^\dagger \Phi_l]
+
\lambda'_{ijkl} tr[\Phi_i^\dagger \Phi_j \Phi_k^\dagger \Phi_l] \nonumber\\
&-&\mu^2 tr[\Delta^\dagger_L \Delta_L+\Delta^\dagger_R \Delta_R] +
\rho_1 [(tr[\Delta^\dagger_L \Delta_L])^2+(tr[\Delta^\dagger_R
\Delta_R])^2] \nonumber\\
&+& \rho_2 (tr[\Delta^\dagger_L \Delta_L \Delta^\dagger_L \Delta_L]
+ tr[\Delta^\dagger_R \Delta_R \Delta_R^\dagger \Delta_R]) + \rho_3
tr[\Delta_L^\dagger \Delta_L \Delta^\dagger_R \Delta_R] \nonumber\\
&+&\alpha_{ij} tr[\Phi_i^\dagger \Phi_j] (tr[\Delta^\dagger_L
\Delta_L] + tr[\Delta^\dagger_R \Delta_R]) \nonumber\\
&+& \beta_{ij} (tr[ \Delta^\dagger_L \Delta_L \Phi_i \Phi_j^\dagger
] + tr[ \Delta^\dagger_R \Delta_R \Phi_i
\Phi_j^\dagger]) \nonumber\\
&+& \gamma_{ij} (tr[ \Delta^\dagger_L \Phi_i \Delta_R
\Phi_j^\dagger] + h.c.),
\end{eqnarray}
where the sums over $i,j,k$ and $l$ run from 1 to 2, with
$\Phi_1=\Phi$ and $\Phi_2=\tilde{\Phi}$. To recover the left-right
symmetry and hermicity condition, the couplings satisfy the
constraints,
\begin{eqnarray}
\mu_{ij} = \mu_{ji}, \, \lambda_{1212} = \lambda_{2121}, \, \lambda_{iijk}=\lambda_{iikj},\nonumber \\
\lambda_{ijkk}=\lambda_{jikk}, \, \lambda'_{ijkl}=\lambda'_{lijk}=\lambda'_{klij}=\lambda'_{jkli}, \nonumber\\
\alpha_{ij} = \alpha_{ji}, \, \beta_{ij} = \beta_{ji}, \, \gamma_{ij} = \gamma_{ji}.
\end{eqnarray}

After the Higgs fields develop
their VEV, we obtain
\begin{eqnarray}\label{VinVev}
V= -\mu^2 (v^2_L + v^2_R) +\frac{\rho}{4} (v^4_L + v^4_R) + \frac{\rho'}{2} v^2_L v^2_R + \frac{\alpha}{2} (v^2_L + v^2_R) v^2 + \gamma v_L v_R v^2,
\end{eqnarray}
where the approximation $v' \ll v$ is used, and the coefficients are
\begin{eqnarray}
&\gamma = 2\gamma_{12},& \nonumber\\
&\alpha = 2 (\alpha_{11} + \alpha_{22} + \beta_{11}),& \nonumber\\
&\rho=4(\rho_1+\rho_2),& \nonumber\\
&\rho'=2 \rho_3.&
\end{eqnarray}
From the minimizing condition $\frac{\partial V}{\partial v_L} =
\frac{\partial V}{\partial v_R} = 0$, if $v_L \neq v_R$, we get the
relations for VEV of Higgs fields,
\begin{eqnarray}\label{vevRelation}
v_L v_R =\frac{\gamma}{\kappa} v^2,
\end{eqnarray}
where $\kappa = \rho- \rho'$. The mass $m_L,m_R$ and $m_D$ will be
of order of $v_L,v_R$ and $v$, respectively. In the next section, we
will classify the types of seesaw mechanism generated from LRSM by
the values of the ratio of Higgs particle self-couplings
$\frac{\gamma}{\kappa}$.

\section{The Seesaw Type and Stability}
We now discuss their contributions to the neutrino masses.
Substituting the relation Eq(\ref{vevRelation}) into the general Type-(I+II) seesaw formula
Eq(\ref{seesaw}), we get
\begin{eqnarray}\label{mN}
m_{\nu}= \left (f (\frac{\gamma}{\kappa}) - \frac{y^2}{f} \right) \frac{v^2}{v_R}.
\end{eqnarray}
According to the formula, following classification can be given.

1) Type-I seesaw: $f (\frac{\gamma}{\kappa}) \ll \frac{y^2}{f}$. It
responds to the case of $m_{\nu} \simeq -\frac{y^2}{f}
\frac{v^2}{v_R} = -m_D m^{-1}_R m^T_D$ dominant, the small neutrino
mass is from the suppression of heavy $v_R$.

2) Type-II seesaw: $f (\frac{\gamma}{\kappa}) \gg \frac{y^2}{f}$. The
term $m_{\nu} \simeq v_L f = m_L$ dominant, while $m_D m^{-1}_R m^T_D$ can be relatively
neglected, i.e. the small neutrino mass is due to the smallness of
$v_L$.

3) Nearly cancellation Type-(I+II) seesaw: $f (\frac{\gamma}{\kappa}) \simeq \frac{y^2}{f}$.
The term $m_L$ and $m_D m^{-1}_R m^T_D$ are comparable in magnitude
and will nearly cancel their contributions to get small neutrino mass, we
will see that this scenario is radiative stable.

However it is classical value at tree level, here we want to
explore the behavior of the $\frac{\gamma}{\kappa}$ defined at the
scale $\mu_0$ under the radiative correction. The correction of
$\gamma$ and $\kappa$ come from the 1-loop correction of the quartic
coupling of operators $\Delta_L \Phi \Delta_R \Phi$ and $\Delta
\Delta \Delta \Delta$, respectively. The renormalization group
equation for $\gamma$ and $\kappa$ take the forms
\begin{eqnarray}\label{RG}
\mu \frac{d \gamma}{d \mu} = \frac{1}{16 \pi^2} [(a_1 \alpha^2 + a_2 \beta^2 + a_3 \gamma^2)
+ (b_1 \alpha + b_2 \beta + b_3 \gamma) y^2 \nonumber\\
+ (c_1 \alpha + c_2 \beta + c_3 \gamma) f^2
+ (d_1 \alpha + d_2 \beta + d_3 \gamma) g^2
+ e_1 g^4 + e_2 f^2 y^2 ], \nonumber\\
\mu \frac{d \kappa}{d \mu} = \frac{1}{16 \pi^2} [(a'_1 \rho_1^2 + a'_2 \rho_2^2 + a'_3 \rho_3^2)
+ (b'_1 \rho_1 + b'_2 \rho_2 + b'_3 \rho_3) f^2 \nonumber\\
+ (c'_1 \rho_1 + c'_2 \rho_2 + c'_3 \rho_3) g^2 + d'_1 g^4 + d'_2
f^4 ],
\end{eqnarray}
in which the coefficients $a,b,c,d,e$ are constants of order $O(1)$
that are determined by computing the corresponding 1-loop Feynman
diagrams. $\alpha_{ij},\beta_{ij},\gamma_{ij},\rho_i$ are coupling
constants in Higgs potential Eq(\ref{v}) and $g$ the gauge coupling.

The Yukawa couplings $f$ and $y$ are of order $O(1)$, but the
typical coupling constants in Higgs potential and the gauge coupling
are generally assumed to be much smaller than that. In fact, for
large couplings, higher order or non-perturbative correction should
be considered and we will not discuss them here. So we assume in this
paper that in Eq(\ref{RG}) they can be approximately dropped, while
only the loops that attribute to Yukawa couplings $f,y$ play dominant role.
We estimate the magnitude of the 1-loop corrections at scale $\mu$
to be
\begin{eqnarray}
\delta \gamma \simeq \frac{-n_{f} f^2 y^2}{16 \pi^2} \ln(\frac{\mu}{\mu_0}), \nonumber\\
\delta \kappa \simeq \frac{-n_{f} f^4}{16 \pi^2}
\ln(\frac{\mu}{\mu_0}),
\end{eqnarray}
where $n_f$ is the number of fermion species. The parameter
$\frac{\gamma}{\kappa}$ is stable only when
\begin{eqnarray}\label{cancel}
0=\delta \left( \frac{\gamma}{\kappa} \right) = \frac{(\delta
\gamma) \kappa - \gamma (\delta \kappa)}{\kappa^2},
\end{eqnarray}
so we get the relation
\begin{eqnarray}
\frac{\gamma}{\kappa} \simeq \frac{y^2}{f^2},
\end{eqnarray}
which is consistent with the nearly cancellation type $f
(\frac{\gamma}{\kappa}) \simeq \frac{y^2}{f}$. In other words, if
$m_{\nu} \simeq 0$ in Eq(\ref{mN}) arises from the cancellation
between $f (\frac{\gamma}{\kappa})$ and $\frac{y^2}{f}$, because of
Eq(\ref{cancel}) it will lead to the stable value of
$\frac{\gamma}{\kappa}$ that suppresses its radiative correction.
Therefore, it is indicated that the scenario of nearly cancellation
Type-(I+II) seesaw is more natural than other types of seesaw when
we consider the factor of their stability. The neutrino mass is
vanished when the cancellation relation $\frac{\gamma}{\kappa} =
\frac{y^2}{f^2}$ is exactly hold as is shown in Eq(\ref{mN}).

The vanishing $m_{\nu}$ can also eliminate another unnaturalness that
the texture of $f$ is not uniquely determined in LRSM \cite{dual}, e.g. if $f$
is allowed, then so is $\hat{f} = \frac{m_{\nu}}{v_L} - f$. We can
see that when $m_{\nu}=0$, $f$ is uniquely determined up to an
unimportant phase or sign.

In this case, the $v_R$ does not need to play the role of
suppressing the neutrino mass, the RHN mass can be scale of
$O(10)$TeV by the constraints of $v_L \lesssim$ 1GeV. This
possibility that $v_R$ can be reachable TeV scale has not been ruled
out by current bounds of experiments \cite{RHNbound}.

\section{Non-Zero Neutrino Masses And Tribimaximal Mixing}
The textures of Yukawa matrices discussed above are simple, in which
the Dirac neutrino masses and the ones coming from the left(right)
Higgs triplet are degenerate,
\begin{eqnarray}\label{md&mlr}
M_D=y v I, \nonumber \\
M_{L(R)}=f v_{L(R)} I.
\end{eqnarray}
The neutrino is massless when the cancellation relation is hold.
However, the masses of neutrino are not trivially vanished. So we
will discuss a deviation from this scenario by perturbations and introducing flavors symmetry to get a more realistic model.

We embed the extra $A_4$ symmetry \cite{A4}
into LRSM by the assignments
\begin{eqnarray}
l_{L(R)}, \, l^c_{L(R)} \sim \underline{\bf{3}}, \, \Phi \sim
\underline{\bf{1}}, \, \Delta_{L(R)} \sim \underline{\bf{1}},
\end{eqnarray}
where, in $A_4$ group, \underline{\bf{3}} stands for the real three-dimensional
irreducible representation and \underline{\bf{1}} for the trivial
one in the three inequivalent one-dimensional representations
\underline{\bf{1}}, \underline{\bf{1}}', \underline{\bf{1}}''. So the invariant Yukawa Lagrangian for their couplings is
\begin{eqnarray}
y (\overline{l_L} l_R)_{\underline{\bf{1}}} \Phi + \tilde{y}
(\overline{l_L} l_R)_{\underline{\bf{1}}} \tilde{\Phi} +
\frac{1}{2} i \tau_2 f \left( (\overline{l_L}
l^c_L)_{\underline{\bf{1}}} \Delta_L + (\overline{l^c_R}
l_R)_{\underline{\bf{1}}} \Delta_R \right) + h.c.,
\end{eqnarray}
in which the tensor product notations and properties of $A_4$ can be found in
Appendix A. Then the above assumptions Eq(\ref{md&mlr})
as well as the lepton mass matrix $M_e=\tilde{y} v I$ can be
achieved automatically, and they preserve the form of Higgs
potential Eq(\ref{v}) since the Higgs fields now are singlets of
$A_4$.

In order to obtain non-trivial mixing, we need to introduce another
scalar $\Sigma \sim \underline{\bf{3}}$ of $A_4$ to generate
off-diagonal elements and assign the gauge group representation
$\Sigma \sim (2,2,0)$ to it. The extra Higgs potential involving
$\Sigma$ and the couplings between $\Sigma$ and $\Phi,\Delta_{L(R)}$
are list in the Appendix B. The extra terms that contribute to the Eq(\ref{VinVev}) have no effect on the relation Eq(\ref{vevRelation}), so the results in the previous sections are still valid.

Now the invariant Lagrangian of
couplings between leptons and $\Sigma$ is written as
\begin{eqnarray}
h (\overline{l}_L l_R)_{\underline{\bf{3}}_s} \cdot \Sigma,
\end{eqnarray}
in which the subscript $\underline{\bf{3}}_s$ denotes the three
dimensional symmetric tensor product as shown in Appendix A. Expanding it into matrix in flavor basis we
obtain the extra contributions
\begin{eqnarray}
\left( \matrix{0 & h v_{\Sigma_3} & h v_{\Sigma_2} \cr h v_{\Sigma_3} & 0 & h v_{\Sigma_1} \cr h v_{\Sigma_2} & h v_{\Sigma_1} & 0} \right),
\end{eqnarray}
where $v_{\Sigma_i} = \langle \Sigma_i \rangle$. In the assumption
of $v_{\Sigma_1}=v_{\Sigma_3}=0$ and $h v_{\Sigma_2} = \delta \neq
0$, the matrix $M_e$ and $M_D$ have similar forms
\begin{eqnarray}\label{md}
M_e(M_D) = \tilde{y}(y) \,\, v I + \left( \matrix{0 & 0 & \delta \cr 0 & 0 & 0 \cr \delta & 0 & 0} \right).
\end{eqnarray}

Now, a deviation of $M_e$ from $M_D$ is needed by perturbations, in
general the vanished elements will have non-zero values $\epsilon$,
and $\delta$ is perturbed to $\delta'$ and $\delta''$,
\begin{eqnarray}
M_e=\left( \matrix{\tilde{y} v & \epsilon_{12} & \delta'' \cr \epsilon_{21} & \tilde{y} v & \epsilon_{23} \cr \delta' & \epsilon_{32} & \tilde{y} v} \right).
\end{eqnarray}
We assume that $\epsilon_{21}, \epsilon_{32} \simeq \delta''$ and
$\epsilon_{12}, \epsilon_{23} \simeq \delta'$, then we get the mass
matrix of charged leptons that can be diagonalized by the unitary
matrix
\begin{eqnarray}
V_e=\frac{1}{\sqrt{3}} \left( \matrix{1 & 1 & 1 \cr 1 & \omega & \omega^2 \cr 1 & \omega^2 & \omega} \right),
\end{eqnarray}
in which $\omega=e^{\frac{2 \pi i}{3}}$, i.e. $V_e^\dagger M_e V_e = diag(m_e, m_{\nu}, m_{\tau})$, where
\begin{eqnarray}
m_e &=& \tilde{y}v + \delta'+\delta'', \nonumber\\
m_{\mu} &=& \tilde{y}v + (\omega \delta' + \omega^2 \delta''), \nonumber\\
m_{\tau} &=& \tilde{y}v + (\omega^2 \delta' + \omega \delta'').
\end{eqnarray}

Under the condition of cancellation relation $\frac{\gamma}{\kappa}
= \frac{y^2}{f^2}$ and non-diagonalized $M_D$ Eq(\ref{md}), a
non-zero neutrino mass matrix now becomes
\begin{eqnarray}
M_{\nu} = \Delta m I - \frac{2f v_L \delta}{y v} \left( \matrix{\frac{\delta}{2 y v} & 0 & 1 \cr 0 & 0 & 0 \cr 1 & 0 & \frac{\delta}{2 y v}} \right),
\end{eqnarray}
where $\Delta m I$ is a perturbation. $M_{\nu}$ can be diagonalized
by the unitary matrix
\begin{eqnarray}
V_{\nu}=\frac{1}{\sqrt{2}} \left( \matrix{1 & 0 & -1 \cr 0 & \sqrt{2} & 0 \cr 1 & 0 & 1} \right).
\end{eqnarray}
we get
\begin{eqnarray}
M^{diag}_{\nu} = V_{\nu}^T M_{\nu} V_{\nu} = diag \left( \Delta m - \frac{2 f v_L \delta}{y v} (1+\frac{\delta}{2 y v}), \Delta m, \Delta m + \frac{2 f v_L \delta}{y v} (1-\frac{\delta}{2 y v}) \right).
\end{eqnarray}
The MNS matrix \cite{MNS} is then obtained as
\begin{eqnarray}
U_{MNS} = V_e^\dagger V_{\nu}= \left( \matrix{\frac{2}{\sqrt{6}} & \frac{1}{\sqrt{3}} & 0 \cr -\frac{\omega}{\sqrt{6}} & \frac{\omega}{\sqrt{3}} & -\frac{e^{i \pi/6}}{\sqrt{2}} \cr -\frac{\omega^2}{\sqrt{6}} & \frac{\omega^2}{\sqrt{3}} & \frac{e^{-i \pi/6}}{\sqrt{2}} } \right),
\end{eqnarray}
which is the tribimaximal mixing matrix up to a phase and hence fits
the neutrino oscillation data well.

\section{Conclusions}
The Type I, II and hybrid (I+II) seesaw mechanisms can be deduced
from the LRSM, and classified by the ranges of the parameter
$\frac{\gamma}{\kappa}$, which represents the ratio of Higgs
particle self-couplings. Assuming that the Yukawa coupling $y,f$ are
of order $O(1)$, then $\frac{\gamma}{\kappa} \ll \left(
\frac{y}{f}\right)^2$ responds to Type-I seesaw,
$\frac{\gamma}{\kappa} \gg \left( \frac{y}{f}\right)^2$ to Type-II
seesaw and $\frac{\gamma}{\kappa} \simeq \left(
\frac{y}{f}\right)^2$ to the comparable or nearly cancellation
Type-(I+II) seesaw. In the limit of weak couplings in Higgs
potential, we find that the parameter region $\frac{\gamma}{\kappa}
\simeq \left( \frac{y}{f}\right)^2 \simeq O(1)$ is more stable
against the radiative correction with respect to other regions,
hence the nearly cancellation Type-(I+II) is more natural than other
types of seesaw in the LRSM.

In the framework of nearly cancellation Type-(I+II) seesaw, the
small neutrino masses arise from the cancellation between the
contribution of the Type-I and Type-II. In this scenario, the RHN
masses can be of order of O(10)TeV and be reachable in future colliders.
We give a realization of this kind of cancellation scenario by
introducing an extra $A_4$ flavor symmetry to govern the textures of
Yukawa coupling matrices. A realistic model that gives non-zero
neutrino masses, charged lepton masses and lepton tribimaximal
mixing is also implemented via introducing perturbations to the textures.

\begin{acknowledgments}
This work was supported in part by the Natural Science Foundation of
China under grant No.90203002.
\end{acknowledgments}

\newpage

\appendix

\section{Basic Properties of $A_4$}
The $A_4$ group has a real three dimensional irreducible representation ${\underline{{\bf 3}}}$, and three inequivalent one dimensional representation ${\underline{{\bf 1}}},{\underline{{\bf 1}}'},{\underline{{\bf 1}}''}$, in which ${\underline{{\bf 1}}}$ stands for the trivial representaion, and ${\underline{{\bf 1}}'}$ and ${\underline{{\bf 1}}''}$ are the non-trivial ones and complex conjugates to each other.

The multiplication rules of their non-trivial tensor products are given as
\begin{eqnarray}
{\underline{{\bf 3}}} \otimes {\underline{{\bf 3}}} = {\underline{{\bf 3}}_s} \oplus {\underline{{\bf 3}}_a} \oplus {\underline{{\bf 1}}} \oplus {\underline{{\bf 1}}'} \oplus {\underline{{\bf 1}}''} \quad {\rm and} \quad {\underline{{\bf 1}}'} \otimes {\underline{{\bf 1}}'} = {\underline{{\bf 1}}''},
\end{eqnarray}
in which the subscript $s(a)$ stands for the symmetric (asymmetric) products. If we set $\psi_i,\phi_i \sim {\underline{{\bf 3}}}$, then
\begin{eqnarray}
({\underline{{\bf 3}}} \otimes {\underline{{\bf 3}}})_{\underline{{\bf 1}}} &=& \psi_1 \phi_1 + \psi_2 \phi_2 + \psi_3 \phi_3, \\
({\underline{{\bf 3}}} \otimes {\underline{{\bf 3}}})_{\underline{{\bf 1}}'} &=& \psi_1 \phi_1 + \omega \psi_2 \phi_2 +\omega^2 \psi_3 \phi_3, \\
 ({\underline{{\bf 3}}} \otimes {\underline{{\bf 3}}})_{\underline{{\bf 1}}''} &=& \psi_1 \phi_1 + \omega^2 \psi_2 \phi_2 +\omega \psi_3 \phi_3, \\
({\underline{{\bf 3}}} \otimes {\underline{{\bf 3}}})_{\underline{{\bf 3}}_s} &=& (\psi_2 \phi_3+\psi_3 \phi_2,  \psi_3 \phi_1+\psi_1 \phi_3, \psi_1 \phi_2+\psi_2 \phi_1),\\
({\underline{{\bf 3}}} \otimes {\underline{{\bf 3}}})_{\underline{{\bf 3}}_a} &=& (\psi_2 \phi_3-\psi_3 \phi_2,  \psi_3 \phi_1-\psi_1 \phi_3, \psi_1 \phi_2-\psi_2 \phi_1),
\end{eqnarray}
with $\omega=e^{\frac{2 \pi i}{3}}$.

\section{Higgs Potential}
In addition to the $A_4$ singlet Higgs fields $\Phi$ and
$\Delta_{L(R)}$, we have introduced another scalar $\Sigma \sim
(2,2,0)(\underline{{\bf3}})$ under the group $G_{LR} \otimes A_4$, so
the extra Higgs potential involving $\Sigma$ should be added. The
potential involving $\Phi$ and $\Delta_{L(R)}$ preserves its form
Eq(\ref{v}) since they are trivial representation of $A_4$, i.e. $\Phi,
\Delta_{L(R)} \sim \underline{{\bf 1}}$, we will not write them here
again. According to the representation assignment of $\Sigma$, the
invariant potential can be written as
\begin{eqnarray}
V(\Sigma)&=&\mu^2_{\Sigma} (\Sigma^\dagger \Sigma)_{\underline{{\bf 1}}}+ \lambda_1^{\Sigma} (\Sigma^\dagger \Sigma)_{\underline{{\bf 1}}} (\Sigma^\dagger \Sigma)_{\underline{{\bf 1}}} + \lambda_2^{\Sigma} (\Sigma^\dagger \Sigma)_{\underline{{\bf 1}}'} (\Sigma^\dagger \Sigma)_{\underline{{\bf 1}}''}\nonumber\\
&+& \lambda_3^{\Sigma} (\Sigma^\dagger \Sigma)_{\underline{{\bf 3}}_s} (\Sigma^\dagger \Sigma)_{\underline{{\bf 3}}_s}+\lambda_4^{\Sigma} (\Sigma^\dagger \Sigma)_{\underline{{\bf 3}}_a} (\Sigma^\dagger \Sigma)_{\underline{{\bf 3}}_a}\nonumber\\
&+&i\lambda_5^{\Sigma} (\Sigma^\dagger \Sigma)_{\underline{{\bf 3}}_s} (\Sigma^\dagger \Sigma)_{\underline{{\bf 3}}_a},\\
V(\Phi, \Sigma)&=&\lambda_{1}^{\Phi \Sigma} (\Sigma^\dagger \Sigma)_{\underline{{\bf 1}}} (\Phi^\dagger \Phi)_{\underline{{\bf 1}}}+\lambda_{2}^{\Phi \Sigma} (\Sigma^\dagger \Phi)_{\underline{{\bf 3}}} (\Phi^\dagger \Sigma)_{\underline{{\bf 3}}} \nonumber\\
&+&\lambda_{3}^{\Phi \Sigma} (\Sigma^\dagger \Phi)_{\underline{{\bf 3}}} (\Sigma^\dagger \Phi)_{\underline{{\bf 3}}}+h.c.,\\
V(\Delta_L,\Delta_R,\Sigma)&=&\lambda_1^{\Delta \Sigma} [ tr(\Delta_L^\dagger \Delta_L)_{\underline{{\bf 1}}}+tr(\Delta_R^\dagger \Delta_R)_{\underline{{\bf 1}}} ] (\Sigma^\dagger \Sigma)_{\underline{{\bf 1}}} \nonumber\\
&+& \lambda_2^{\Delta \Sigma} \Sigma^\dagger_{\underline{{\bf 3}}} \left( [\Delta_L, \Delta_L^\dagger]_{\underline{{\bf 1}}} + [\Delta_R, \Delta_R^\dagger]_{\underline{{\bf 1}}} \right) \Sigma_{\underline{{\bf 3}}}\label{deltaSigma}.
\end{eqnarray}
There is no renormalizable term simultaneously involving $\Phi,\Delta_{L(R)}$ and $\Sigma$,
\begin{eqnarray}
V(\Phi,\Delta_L,\Delta_R,\Sigma)=0.
\end{eqnarray}
So the total Higgs potential is given by
\begin{eqnarray}
V=V(\Phi,\Delta_L,\Delta_R)+V(\Sigma)+V(\Phi, \Sigma)+V(\Delta_L,\Delta_R,\Sigma).
\end{eqnarray}

\end{document}